\documentclass[runningheads]{llncs}
\usepackage[T1]{fontenc}
\usepackage{graphicx}
\usepackage{mathtools}
\usepackage{amssymb}
\usepackage{subcaption}
\usepackage{tabularx}
\usepackage{booktabs}
\usepackage{float}
\usepackage{array}
\usepackage{comment}
\usepackage{hyperref}
\usepackage{multirow}
\usepackage{marvosym}
\usepackage[printonlyused,withpage]{acronym}

\acrodef{STFT}{short-time Fourier transform}
\acrodef{ESTOI}{extended short-time objective intelligibility}
\acrodef{DNN}{deep neural network}
\acrodef{VAE}{variational auto-encoder}
\acrodefplural{VAEs}{variational auto-encoders}
\acrodef{EM}{expectation-maximisation}
\acrodef{TF}{time-frequency}
\acrodef{ELBO}{evidence lower bound}
\acrodef{SDR}{signal-to-distortion ratio}
\acrodef{PESQ}{perceptual evaluation of speech quality}
\acrodef{SNR}{signal-to-noise ratio}
\acrodef{DNNs}{deep neural networks}
\acrodef{SDE}{stochastic differential equation}
\acrodef{GAN}{generative adversarial networks}
\acrodefplural{GANs}{generative adversarial networks}
\acrodef{SI-SDR}{scale-invariant signal-to-distortion ratio}
\acrodef{MOS}{mean opinion score}
\acrodef{SGMSE+}{score-based generative model for speech enhancement}
\acrodef{NCSNPP++}{Noise-Conditional Score Network}
\acrodef{WSJ}{Wall Street Journal}
\acrodef{PC}{Predictor-Corrector}
\acrodef{USS}{universal source separation}
\acrodef{DDPM}{denosing diffusion probabilistic model}
\acrodef{SGM}{score-based generative model}
\acrodef{SED}{sound event detection}

\begin{document}
\title{Extract and Diffuse: Latent Integration for Improved Diffusion-based Speech and Vocal Enhancement}

\renewcommand{\thefootnote}{}
\footnotetext{$^*$ Equal contribution; \ \ \Letter: Corresponding author.}
\renewcommand{\thefootnote}{\arabic{footnote}} 
\titlerunning{Extract and Diffuse}
%
\author{Yudong Yang\inst{1*} \and
Zhan Liu\inst{1*} \and
Wenyi Yu\inst{1} \and
Guangzhi Sun\inst{2} \and \\
Qiuqiang Kong\inst{3} \and Chao Zhang\inst{1}\textsuperscript{\Letter}\vspace{2mm}}
\authorrunning{Y. Yang, Z. Liu et al.}
%
\institute{Tsinghua University \and
University of Cambridge \and Chinese University of Hong Kong \\
\email{\{yang-yd21, liuzhan22\}@mails.tsinghua.edu.cn}\\}
\maketitle              
\begin{abstract}
Diffusion-based generative models have recently achieved remarkable results in speech and vocal enhancement due to their ability to model complex speech data distributions. While these models generalize well to unseen acoustic environments, they may not achieve the same level of fidelity as the discriminative models specifically trained to enhance particular acoustic conditions. In this paper, we propose \textbf{Ex-Diff}, a novel score-based diffusion model that integrates the latent representations produced by a discriminative model to improve speech and vocal enhancement, which combines the strengths of both generative and discriminative models. Experimental results on the widely used MUSDB dataset show relative improvements of 3.7\% in SI-SDR and 10.0\% in SI-SIR for vocal enhancement tasks compared to the baseline diffusion model. Additionally, case studies are provided to further illustrate and analyze the complementary nature of generative and discriminative models in this context.

\end{abstract}
\section{Introduction}

Speech and vocal enhancement involve isolating clean speech or singing voices from audio recordings that contain acoustic noise\cite{hendriks2022dft} or musical accompaniments and mixes \cite{defossez2019music}. Traditional signal processing approaches often rely on assumptions about the structure of spectrograms to differentiate between speech (or vocals) and noise (or accompaniments) \cite{roweis2000one,gerkmann2015phase}. In contrast, deep learning methods can effectively learn these underlying relationships from large datasets, enabling more powerful speech and vocal enhancement\cite{fonseca2021selfsupervised,helwani2024soundsourceseparationusing,zhang2023towarduniversal,koizumi2020speechenhancement}. Deep learning models can be broadly categorized into discriminative and generative models, depending on whether they aim to model posterior distributions or prior/likelihood distributions.

\textit{Discriminative models} apply learnable regression functions to map noisy speech to clean speech, typically using either time or frequency domain methods \cite{kong2023universal,defossez2019demucs,stoller2018wave,jansson2017singing,wang2023framework,tzinis2020sudo,chao2024mamba}. In contrast, \textit{generative models} \cite{song2020score,ho2020denoising,song2019generative,goodfellow2014generative,rezende2015variational,shi2024ensemble,li2024diffusionbasedgenerativemodel,jukic2024schr,wang2024diffusion} focus on learning the underlying statistical properties of clean speech data distributions (\textit{e.g.}, spectral characteristics and temporal dynamics) to allow them to perform better in mismatched training and test conditions \cite{fang2021variational,kim2021multi,welker2022speech}. Among generative models, diffusion models have gained prominence for their competitive and complementary performance compared to the discriminative models \cite{welker2022speech,richter2023speech,sepanddiff}. However, due to their generative nature, diffusion models may introduce extraneous sounds in regions where no speech or vocals are present.

In this paper, we propose \textbf{Ex}tract and \textbf{Diff}use ({Ex-Diff}), a novel approach that combines the strengths of both generative and discriminative methods for improved diffusion-based speech and vocal enhancement. {Ex-Diff} leverages pre-trained latent representations from the \ac{USS} model \cite{kong2023universal}, a discriminative model combining \ac{SED} with a conditional source separation model, as conditioning inputs to a score-based diffusion model via a cross-attention mechanism. Unlike the approach in \cite{richter2023speech}, which conditions directly on the audio mixture, using latent representations provides a clearer indication of what to extract and enhance, thus reducing the likelihood of generating unexpected sounds while benefiting from the improved quality of generative modeling.
We evaluate {Ex-Diff} on the VoiceBank-DEMAND \cite{thiemann2013diverse} and MUSDB18 \cite{rafii2017musdb18} datasets. The results show {Ex-Diff} achieves similar SI-SDR on VoiceBank-DEMAND and a 3.7\% relative improvement in SI-SDR on MUSDB18 compared to the baseline diffusion model \cite{richter2023speech}.

\section{Related Work}

\subsection{Discriminative Models for Speech and Vocal Enhancement}
Discriminative models are trained to map noisy speech to a clean target. Time domain models such as Demucs \cite{defossez2019demucs} and Wave-U-Net \cite{stoller2018wave} directly estimate the final separation target by an encoder-decoder architecture. Frequency-domain models leverage a spectrogram, such as \ac{STFT}, to fully utilize harmonic features patterns \cite{jansson2017singing}. Kong et al. \cite{kong2023universal} proposed a \ac{USS} method that uses a \ac{SED} model as a query network to estimate the probabilities of vocal occurrence at different locations. The output embeddings from the query network are then applied to an encoder-decoder discriminative model, ResUNet \cite{kong2021decoupling}, to perform source separation on weakly labeled data. The method showed the ability to separate musical sources on MUSDB18 dataset. USS did not generalize well to unseen song conditions, including different languages, musical styles, and \ac{SNR} levels.

\subsection{Diffusion Models for Speech and Vocal Enhancement}
Ho et al. \cite{ho2020denoising} proposed the idea of generating high-quality samples by \textit{diffusion (probabilistic) model}. The idea behind the {diffusion model} is to perturb the clean data with multiple scales of Gaussian noise \cite{ho2020denoising,song2020score}, transferring the data to a Gaussian distribution, and then learning a score model \cite{song2019generative} to reverse the disturbing process. 
Welker et al. perturbed the clean speech with both Gaussian noise and environmental noise derived from the difference between clean and noisy speech in the forward process by introducing a drift term in the forward \ac{SDE} \cite{welker2022speech}. 
Plaja et al. proposed a training and sampling strategy for singing voice extraction using \ac{DDPM} \cite{plaja2022diffusion}, which generalizes well to unseen data. However, its \ac{SDR} results on the MUSDB18 test set (5.59 dB) indicate a lower separation performance compared to USS (8.12 dB).

\section{Diffusion for Speech Enhancement}

For speech enhancement, we define \textit{clean speech} as audio containing only the speaker's voice, without any additional sources. Speech enhancement, from the perspective of generative models, treats the given noisy speech as a condition, and then samples the corresponding clean speech from the distribution of clean speech signals. Following the baseline approach \cite{richter2023speech}, {Ex-Diff} adopts a diffusion-based method to incorporate the noisy speech condition into both the diffusion forward process and the reverse process.

\subsubsection{{Forward Process}}

The forward process involves gradually adding Gaussian noise to clean speech. Following Song et al. \cite{song2020score}, we design a stochastic diffusion process $\{\mathbf x_t\}_{t=0}^T$ that is the solution of the following linear \ac{SDE}:
\begin{equation}
\label{forward}
    \mathrm{d}\mathbf{x}_{t} =f\left (\mathbf{x}_{t}  \right )\mathrm{d}t+g\left ( t \right ) \mathrm{d}\mathbf{w},
\end{equation}
where $\mathbf{x}_{t}$ is the current state speech representation, $t\in [0, T]$ is a continuous time-step variable, $\mathbf{w}$ is the standard Wiener process, with $\mathbf{x}_{0}$ representing clean speech, $\mathbf{x}_{T}$ representing Gaussian distribution centered around the noisy speech. We incorporate the noisy speech $\mathbf{y}$ into the \ac{SDE} by modifying the \textit{drift coefficient} to $f\left (\mathbf{x}_{t}, \mathbf{y} \right) \coloneqq \gamma \left (\mathbf{y} - \mathbf{x}_{t} \right)$
where $\gamma$ is a constant controlling the transition from $\mathbf{x}_{0}$ to $\mathbf{y}$. The diffusion coefficient $g\left ( t \right )$ is defined as:\newline
\begin{equation}
\label{gt}
    g\left ( t \right ) \coloneqq \sigma_{\mathrm{min}}\left ({\sigma _{\mathrm{max}}}/{\sigma_{\mathrm{min}}}  \right ) ^{t} \sqrt{2\mathrm{log}\left ({\sigma _{\mathrm{max}}}/{\sigma_{\mathrm{min}}}  \right ) },
\end{equation}
where $\sigma_{\mathrm{min}}$ and $\sigma_{\mathrm{max}}$ are parameters defining the
noise schedule of the Wiener process.\newline

\subsubsection{{Reverse Process}}

The reverse process is used to estimate the clean speech by reversing the forward process. This is achieved by solving the following differential equation \cite{anderson1982reverse}:
\begin{equation}
\begin{aligned}
\label{reverse}
\mathrm{d}\mathbf{x}_{t}=\left[g\left(t\right)^{2}\nabla_{\mathbf{x}_{t}}\log
p_{t}\left(\mathbf{x}_{t}\mid \mathbf{y}\right)-f\left(\mathbf{x}_{t},\mathbf{y}\right)\right]\mathrm{d}t+g\left(t\right)\mathrm{d}\bar{\mathbf{w}}
\end{aligned}
\end{equation}
where $\bar{\mathbf{w}}$ is a standard Wiener process running backwards in time. The \textit{score function} $\nabla_{\mathbf{x}_{t}}\log p_{t}\left(\mathbf{x}_{t}\mid \mathbf{y}\right)$ is approximated by a deep learning model termed as \textit{score model}, which is denoted as $s_{\theta}\left(\mathbf{x}_{t}, \mathbf{y}, t\right)$.
Now, we can sample $\mathbf{x}_{T} \sim \mathcal{N_{\mathbb{C}}} \left(\mathbf{x}_{T};\mathbf{y},\sigma (T)^2\mathrm{I}\right)$ where $\mathbf{x}_{T}$ is a strongly corrupted data distribution of noisy speech $\mathbf{y}$.
Once the score model is trained, the reverse \ac{SDE} defined by Eqn.~\eqref{reverse} can be solved using a Predictor-Corrector sampling procedure iteratively to estimate the clean speech \cite{song2020score}.

\subsubsection{{Training Objective}}

This section discusses the objective function used to train $s_{\theta} (\mathbf{x}_{t},\mathbf{y}, t)$.
This section discusses the objective function used to train $s_{\theta} (\mathbf{x}_{t},\mathbf{y}, t)$.
Since Eqn.~\eqref{forward} defines a Gaussian process, the mean and variance of step $\mathbf{x}_{t}$ are easily determined given the initial state $\mathbf{x}_{0}$ and condition $\mathbf{y}$ \cite{sarkka2019applied}:
\begin{equation}
    p(\mathbf{x}_{t}\mid \mathbf{x}_{0},\mathbf{y})=\mathcal{N_{\mathbb{C}}}(\mathbf{x}_{t};\mu (\mathbf{x}_{0}, \mathbf{y}, t),\sigma(t)^2\mathrm{\mathbf{I}})
\end{equation}
where
\begin{equation}
\label{miu}
    \mu (\mathbf{x}_{0},\mathbf{y},t)=\mathrm{e}^{-\gamma t} \mathbf{x}_{0}+(1-\mathrm{e}^{-\gamma t})\,\mathbf{y}
\end{equation}
and
\begin{equation*}
    \sigma(t)^2 = \frac{
        \sigma_{\mathrm{min}}^2 \left( \left( \sigma_{\mathrm{max}} / \sigma_{\mathrm{min}} \right)^{2t} - \mathrm{e}^{-2\gamma t} \right) \log\left( \sigma_{\mathrm{max}} / \sigma_{\mathrm{min}} \right)
    }{\gamma + \log\left( \sigma_{\mathrm{max}} / \sigma_{\mathrm{min}} \right)}.
    \
\end{equation*}
Then $\mathbf{x}_{t}$ can be efficiently computed as $    \mathbf{x}_{t} = \mu(\mathbf{x}_{0}, \mathbf{y}, t)+\sigma(t)\mathrm{\mathbf{z}}
$, 
where $\mathrm{\mathbf{z}}\sim \mathcal{N_{\mathbb{C}}}(\mathrm{\mathbf{z}};\mathbf{0},\mathrm{\mathbf{I}})$.

Using the \textit{denoising score matching} principle \cite{vincent2011connection}, the score of the perturbation kernel $\nabla_{\mathbf{x}_{t}}\log p(\mathbf{x}_{t}\mid \mathbf{x}_{0}, \mathbf{y})$ simplifies to:
\begin{equation}
    \nabla_{\mathbf{x}_{t}}\log p(\mathbf{x}_{t}\mid \mathbf{x}_{0}, \mathbf{y})=-{\mathrm{\mathbf{z}}}/{\sigma(t)}. 
\end{equation}

After inputting $(\mathbf{x}_{t}, \mathbf{y}, t)$ into the score model, the final loss is expressed as an unweighted $L_{2}$ loss between the model's output and the score of the perturbation kernel. The training objective is then given by:
\begin{equation}
\arg\min_{\theta} \mathbb{E}_{t, (\mathbf{x}_0, \mathbf{y}), \mathrm{\mathbf{z}}, \mathbf{x}_t \mid (\mathbf{x}_0, \mathbf{y})} \left[
  \left\| s_{\theta}(\mathbf{x}_t, \mathbf{y}, t) + {\mathrm{\mathbf{z}}}/{\sigma(t)} \right\|^{2}_2
\right].
\end{equation}

\section{Extract and Diffuse}

In the vocal enhancement task which isolates the human voice from a musical background, the voice does not always span the entire musical segment. However, due to the generative nature of the diffusion model, the enhanced signal may introduce sounds in areas where neither speech nor vocals are present. Therefore, it is crucial to have a clear indication of what should be extracted and enhanced within an audio segment. In this paper, we propose a novel method to address this issue. First, a pre-trained discriminative model is used to extract the latent representation of the vocals in the audio, denoted as $\mathbf{l}$. Next, this latent representation is incorporated into the score model calculation by $s_{\theta} = s_{\theta}(\mathbf{x}_{t},\mathbf{y},t,\mathbf{l})$, providing a clear indication of what to enhance during the diffusion process. The details of the latent representation extraction and diffusion procedures are presented in this section.

\subsection{Latent Representation Extraction}

Following the procedure in USS \cite{kong2023universal}, a pre-trained audio neural networks (PANNs) model \cite{kong2020panns} is used to extract the latent representation of vocals in the input audio.

The PANNs model is trained with weakly labeled data from AudioSet \cite{gemmeke2017audio}, and it can localize the occurrence of sound classes as a SED system. For the audio clip, the SED system can produce a frame-wise event prediction $p_\text{SED} \in [0,1]^{T \times K}$, where $T$ is the number of frames and $K$ is the predefined number of sound classes. In our model, each raw audio clip is first divided into segments of 64{,}000 samples. For the $i$-th segment, the PANNs model yields penultimate-layer features of size $(T,H)$, which are averaged over the temporal dimension to produce a segment-level vector $\mathbf{l}_i = \tfrac{1}{T}\sum\nolimits_{t=1}^{T}\mathbf{h}_t^{(i)}$, where $\mathbf{h}_t^{(i)} \in \mathbb{R}^H$ denotes the feature of the $t$-th frame in segment $i$. The segment-level vectors are then concatenated in temporal order to form the final latent representation $\mathbf{l}$ with shape $(L,H)$, where $L$ is the number of segments and the default dimension is $H{=}2048$. To preserve temporal dependencies, positional encodings are added, and cross-attention is applied along the temporal dimension.

We redefine the score model as $s_{\theta} = s_{\theta}(\mathbf{x}_{t},\mathbf{y},t,\mathbf{l})$, and the reverse \ac{SDE} in Eqn.~\eqref{reverse} can be expressed as:
\begin{equation}
    \mathrm{d}\mathbf{x}_{t}=\left[-f\left(\mathbf{x}_{t},\mathbf{y}\right)+g\left(t\right)^{2}s_{\theta}\left(\mathbf{x}_{t},\mathbf{y},t, \mathbf{l}\right)\right]+g\left(t\right)\mathrm{d}\bar{\mathbf{w}}.
\end{equation}
The loss function can be expressed as:
\begin{equation*}
    \arg\min_{\theta} \mathbb{E}_{t, (\mathbf{x}_0, \mathbf{y}), \mathrm{z},\mathbf{x}_t \mid (\mathbf{x}_0, \mathbf{y})} \left[
  \left\| s_{\theta}(\mathbf{x}_t,\mathbf{y}, t, \mathbf{l}) + {\mathrm{\mathbf{z}}}/{\sigma(t)} \right\|^{2}_2
\right].
\end{equation*}

\subsection{The Conditional Reverse Diffusion Process in Ex-Diff}

\begin{figure}[t]  
    \centering
    \hspace{0cm}
    \includegraphics[width=0.99\textwidth]{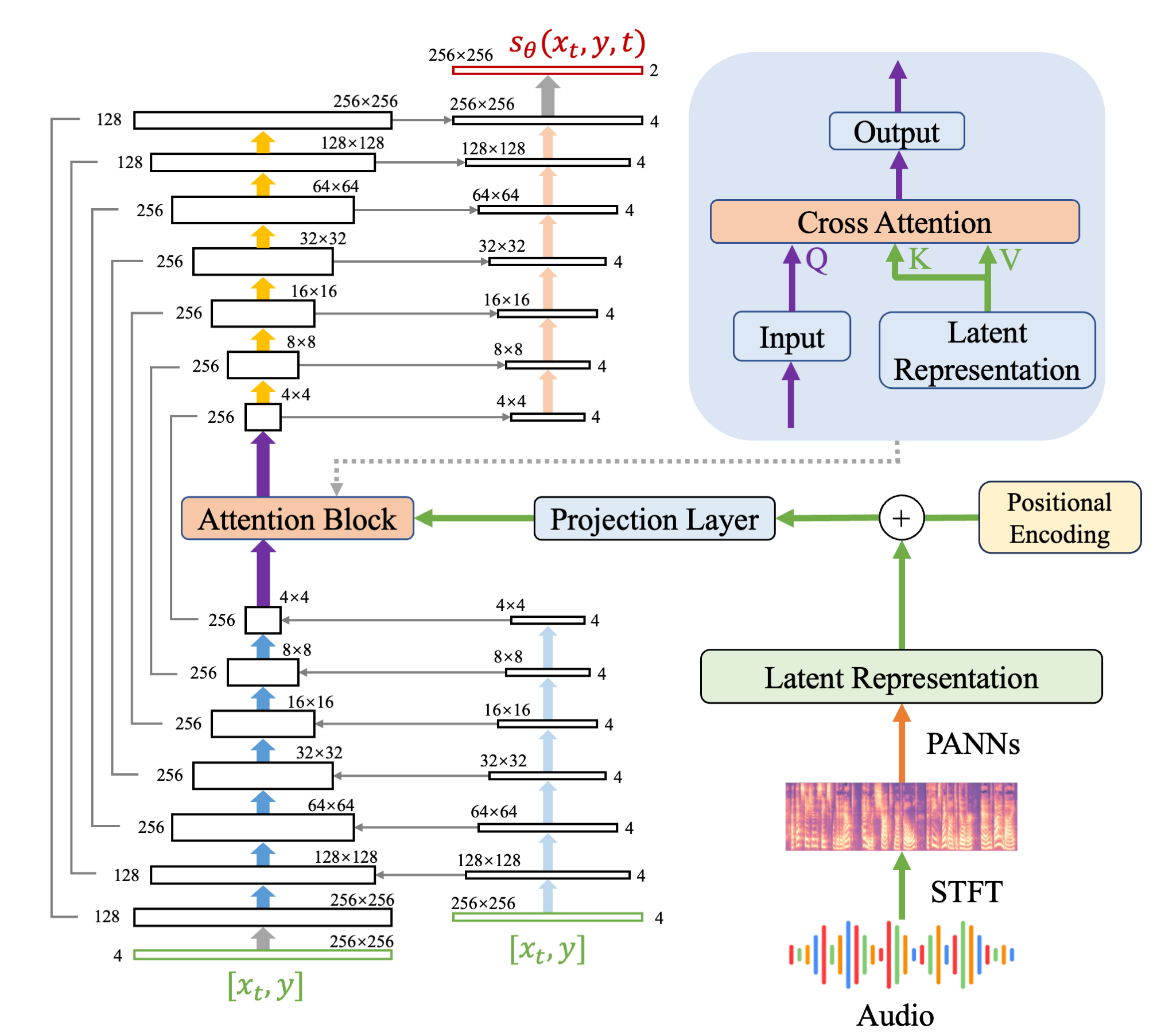}  
    \caption{The multi-resolution U-Net used as the score model. Latent representations are integrated through cross-attention.}
    \label{structure}
\end{figure}

The noise conditional score network \cite{song2020score} is used as the backbone of our score model. As shown in Fig.~\ref{structure}, our model has a multi-resolution U-Net structure \cite{richter2023speech}. Progressive growth of the input \cite{karras2020analyzing} is also used on the basis where the upsampling and downsampling layers use the residual network blocks taken from the BigGAN structure \cite{brock2018large}. For each set of speech data, the corresponding $\mathbf{x}_{t}$ and $\mathbf{y}$ are concatenated at the input of the U-Net and fed into the model. Meanwhile, the noisy speech spectrogram is input into a pre-trained PANNs model to generate the latent representation. Since the latent representation captures the audio events occurring in each frame of the original input audio and has strong temporal characteristics, positional encoding \cite{Vaswani2017AttentionIA} is applied to it, allowing our model to learn important temporal information. Subsequently, the latent representation is passed through a linear layer to be reshaped to match the dimensions of the feature map in the U-Net bottleneck layer. This reshaped latent representation is then fed into the bottleneck layer of the U-Net via cross-attention \cite{Vaswani2017AttentionIA} as shown in Fig.~\ref{structure}, where $\mathbf{Q}$ is derived from the feature map of the U-Net bottleneck layer, $\mathbf{K}$ and $\mathbf{V}$ are produced from the latent representation. This process generates a new feature map that continues to participate in the subsequent U-Net computations.

\section{Experimental Setup}

\subsection{Datasets and Evaluation Metrics}

Regarding speech enhancement, the VoiceBank-DEMAND dataset is used, while the MUSDB18 dataset is used regarding vocal enhancement. PESQ \cite{rix2001perceptual}, ESTOI \cite{jensen2016algorithm}, SI-SDR, SI-SIR, SI-SAR \cite{le_roux2019sdr} are used as the evaluation metrics.

\begin{table*}[t]
\caption{Left: Speech enhancement results obtained for VoiceBank-DEMAND test set. 
Right: Vocal enhancement results obtained for the mix3dB-MUSDB test set. 
Note that models trained on VB or AS+VB are evaluated only on the speech enhancement task, 
while models trained on M3 are evaluated only on the vocal enhancement task. 
Each result is presented in the form of mean $\pm$ standard deviation. 
AS, M3 and VB refer to the AudioSet, MUSDB-mix3dB and VB-DMD datasets, respectively.}
\setlength{\tabcolsep}{3pt}
\label{bigtable}
\vspace{0.2cm}
\renewcommand{\arraystretch}{1.4}

\resizebox{\textwidth}{!}{
\begin{tabular}{lcccccc|cccccc}
\toprule
\multicolumn{1}{c}{\multirow{2}{*}{Method}} &
  \multirow{2}{*}{Training Set} &
  \multicolumn{5}{c|}{Speech Enhancement} &
  \multicolumn{5}{c}{Vocal Enhancement}\\ 
  \cline{3-12} 
\multicolumn{1}{c}{}  &                                              & PESQ   & ESTOI  & SI-SDR [dB] & SI-SIR [dB]  & SI-SAR [dB] & PESQ  & ESTOI  & SI-SDR [dB] & SI-SIR [dB]  & SI-SAR [dB] \\ 
\midrule
Mixture & - & $1.97 \pm 0.75 $ & $0.79\pm0.15$ & $8.4\pm5.6$ & $8.5\pm5.6$ & $47.5\pm10.4$ & $1.12\pm0.07$ & $0.60\pm0.10$ & $3.0\pm0.0$ & $3.0\pm0.0$ & $50.5\pm10.0$ \\
\midrule
SGMSE\cite{richter2023speech}                    & VB / M3                            & $\textbf{2.91} \pm \textbf{0.63}$  &$0.86 \pm 0.10$  &$17.4 \pm 3.3$  &$28.6 \pm 5.4$  &$18.0 \pm 3.3$  &$1.95 \pm 0.50$ & $\textbf{0.70} \pm \textbf{0.13}$  &$10.7 \pm 3.5$ & $23.9 \pm 6.9$ & $11.1 \pm 3.3$  \\
USS \cite{kong2023universal}               & AS                          & $2.19 \pm 0.56$  & $0.79\pm0.14$ & $16.8\pm4.8$ & $31.0\pm11.6$ & $17.8\pm4.7$ &$1.25\pm0.31$ &$0.29\pm0.20$ &$-2.5\pm6.0$ & $28.8\pm8.9$ & $-2.5\pm6.0$  \\
USS+SGMSE& AS+VB / M3 & $2.59\pm0.61$ & $0.83\pm0.12$ & $\textbf{17.9}\pm\textbf{4.7}$ & $\textbf{37.4}\pm\textbf{9.9}$ & $\textbf{18.1}\pm\textbf{4.7}$ & $1.24\pm0.35$ & $0.25\pm0.20$ & $-3.8\pm5.7$ & $\textbf{29.5}\pm\textbf{8.2}$ & $-3.8\pm5.7$ \\
Ex-Diff    & VB / M3                              &$2.84\pm0.60$  &$\textbf{0.87}\pm\textbf{0.10}$ &$17.4\pm3.1$ &$30.6\pm6.5$ &$17.7\pm3.1$  &$\textbf{2.01} \pm \textbf{0.53}$  &$\textbf{0.70} \pm \textbf{0.14}$ &$\textbf{11.1} \pm \textbf{3.7}$  &$26.3 \pm 6.1$    &$\textbf{11.3} \pm \textbf{3.6}$       \\
\bottomrule
\end{tabular}}
\vspace{0cm}

\vspace{-0.5cm}
\label{tab:exp-t2a-retrieval}
\end{table*}

\subsection{Training Details}

The hyperparameters in Eqns.~\eqref{gt} and~\eqref{miu} are set to $\sigma_{\mathrm{min}}=0.05$, $\sigma_{\mathrm{max}}=0.5$, and $\gamma=1.5$. All audio is resampled to 16kHz, and a learning rate of $10^{-4}$ with the Adam optimizer is used with a mini-batch size of 8. An exponential moving average to the model weights with a decay rate of $0.999$. During the inference process, the number of reverse steps is set to $N=40$ for speech enhancement and $N=90$ for vocal enhancement, as this configuration yielded the best performance under our architecture based on the experimental results. The other sampling parameters are kept consistent with the baseline model \cite{richter2023speech}. 

\section{Experimental Results}

\subsection{Speech Enhancement Results}

The left part of Table~\ref{bigtable} presents the speech enhancement results on the VoiceBank-DEMAND test set. Our proposed Ex-Diff model shows strong performance across a range of objective metrics, showcasing significant improvements over USS in PESQ, ESTOI, and SI-SDR, achieving similar performance in SGMSE, and surpassing it in the SI-SIR metric. However, it is worth noting that the combined USS+SGMSE model exhibits a marginally better SI-SDR score compared to Ex-Diff, which can be attributed to two key factors. First, USS as a discriminative model, uses an L1 loss function directly calculated on the predicted and ground truth waveforms, inherently favouring higher SI-SDR scores due to the better (compared to the diffusion training loss) correlation between minimizing waveform distance and maximizing signal-to-interference ratio \cite{lemercier2023diffusion,bralios2023latent}. Second, the subsequent refinement using the SGMSE diffusion model further contributes to this bias by directly targeting waveform similarity, a factor heavily reflected in SI-SDR. Conversely, Ex-Diff, as an end-to-end generative model, focuses on learning the underlying speech distribution to produce enhanced speech with high perceptual quality. While this approach may not prioritize SI-SDR optimization to the same extent as a concatenated discriminative-diffusion approach, it avoids potential drawbacks of model cascading, such as error propagation and increased complexity. Furthermore, it is crucial to recognize that SI-SDR alone cannot fully encapsulate the nuanced aspects of speech quality, such as naturalness and listener preference, areas where generative approaches like Ex-Diff often excel.

\subsection{Vocal Enhancement Results}

The MUSDB18 dataset is used for vocal enhancement. According to \cite{opensourceseparation:book}, training models with an appropriate SNR and using \textit{incoherent mixing} for data augmentation can lead to better performance.

We reprocessed the \ac{SNR} of the MUSDB18 dataset to 3dB, and adopted an \textit{incoherent mixing} strategy for the training set, which enhances both the quantity and quality of the data.

The experimental results on MUSDB-mix3dB test set, as shown in Table~\ref{bigtable}, indicate that our model outperforms all baseline models. It should be noted that, to ensure a fair comparison, our baseline models were also trained from scratch on the MUSDB-mix3dB dataset. In which the USS model achieves good results on the original MUSDB dataset, but performs poorly on the processed MUSDB dataset, which also reflects that the generative model has stronger generalization ability. \footnote{The demo can be accessed at: \href{https://liuzhan22.github.io/ex-diff-demo/}{https://liuzhan22.github.io/ex-diff-demo/}.}.

\subsection{Discussions on Selected Examples}

\begin{figure}[t]
    \centering
    
    \begin{subfigure}{0.49\linewidth}
        \centering
        \includegraphics[width=\linewidth]{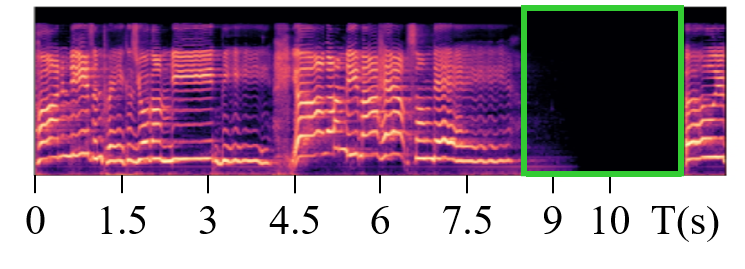}
        \vspace{-20pt}
        \caption{Ground truth}
        \label{fig:image3}
    \end{subfigure}
    \hfill
    \begin{subfigure}{0.49\linewidth}
        \centering
        \includegraphics[width=\linewidth]{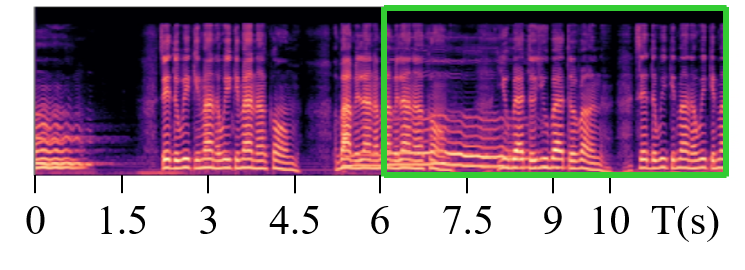}
        \vspace{-20pt}
        \caption{Ground truth}
        \label{fig:image4}
    \end{subfigure}
    
    \begin{subfigure}{0.49\linewidth}
        \centering
        \includegraphics[width=\linewidth]{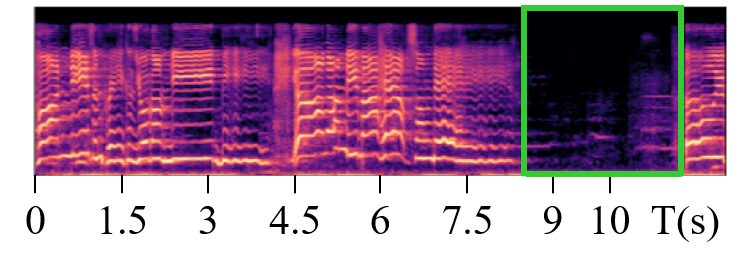}
        \vspace{-20pt}
        \caption{SGMSE enhanced}
        \label{fig:image3}
    \end{subfigure}
    \hfill
    \begin{subfigure}{0.49\linewidth}
        \centering
        \includegraphics[width=\linewidth]{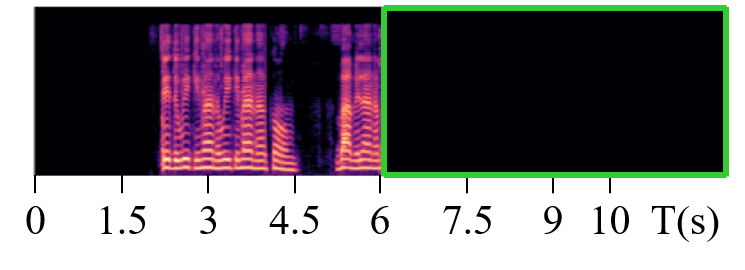}
        \vspace{-20pt}
        \caption{USS enhanced}
        \label{fig:image4}
    \end{subfigure}
    
    \begin{subfigure}{0.49\linewidth}
        \centering
        \includegraphics[width=\linewidth]{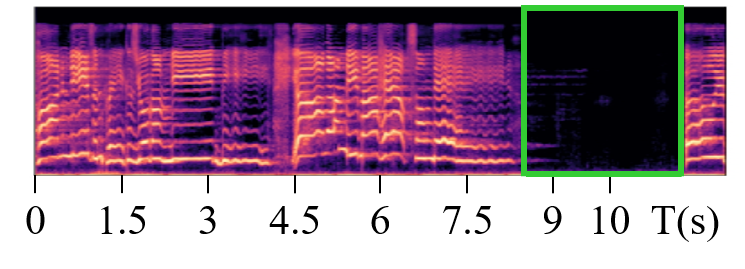}
        \vspace{-20pt}
        \caption{Ex-Diff enhanced}
        \label{fig:image3}
    \end{subfigure}
    \hfill
    \begin{subfigure}{0.49\linewidth}
        \centering
        \includegraphics[width=\linewidth]{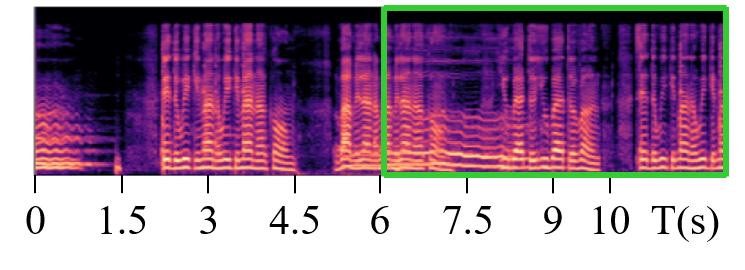}
        \vspace{-20pt}
        \caption{Ex-Diff enhanced}
        \label{fig:image4}
    \end{subfigure}
    
    \caption{The Mel-spectrograms of vocal enhancement results using different models on two speech samples (left and right).}
    \label{mel}
\end{figure}

The Mel-spectrograms of two vocal enhancement test set examples are selected as shown in the left and right parts of Fig.~\ref{mel} respectively. The analysis of the Mel-spectrograms reveals that the vocals enhanced by SGMSE \cite{richter2023speech} introduce sounds in parts where there should be neither speech nor vocals. On the other hand, USS \cite{kong2023universal} executes the separation task excessively, resulting in enhanced speech that contains neither vocals nor accompaniment, producing blank audio segments.

\subsection{Ablation Study}

Ablation studies have been performed to find a better latent representation fusion structure. A total number of four fusion architectures is implemented. \textbf{1)} Integrating latent representations into the score model via cross-attention at the bottleneck of U-Net. \textbf{2)} Integrating the latent representation using cross-attention at 3 layers, corresponding to the upsampling stage, the bottleneck layer, and the downsampling stage of the U-Net. \textbf{3)} Fusing the latent representation using a transformer-like attention block at the bottleneck layer of the U-Net. \textbf{4)} Concatenating the latent representation with $\mathbf{x}$ and $\mathbf{y}$ at the input of the score model, and feeding the combined $(\mathbf{x}, \mathbf{y}, \mathbf{l})$ into the score model to participate in the computation throughout the entire U-Net. The results indicate that using cross-attention only once at the bottleneck layer of the U-Net performs the best, as this connection maximizes the use of the information from the latent representations.

\begin{table}[H]
\caption{Results of the ablation study. A single cross-attention layer (as in Fig. \ref{structure}) performs the best. Removing the attention mechanism by concatenating the latent representation with $\mathbf{x}$ and $\mathbf{y}$, using a full Transformer-like block, or adding up to three attention blocks did not yield better results.}
\footnotesize
\setlength{\tabcolsep}{3pt}
\centering
\begin{tabularx}{1.0\linewidth}{@{\hspace{0cm}}Xcccc@{\hspace{0cm}}}
\toprule
 & PESQ & ESTOI & SI-SDR & SI-SIR \\
\midrule
Mixture & $1.12$ & $0.60$ & $3.0$ & $3.0$ \\
\midrule
\textbf{1x Cross-attention block} & $\mathbf{2.01}$ & $0.70$ & $\mathbf{11.1}$ & $\mathbf{26.3}$ \\
Concatenate & $1.91$ & $0.71$ & $10.7$ & $22.5$ \\
SGMSE \cite{richter2023speech} & $1.95$ & $0.70$ & $10.7$ & $23.9$ \\
1x Transformer-like attn. block & $1.52$ & $0.67$ & $8.3$ & $15.7$ \\
3x Cross-attention blocks & $1.66$ & $0.69$ & $9.0$ & $17.6$ \\
\bottomrule
\end{tabularx}

\vspace{-2em}
\label{tab:speaking_style}
\end{table}

\section{Conclusion}
In this paper, we propose a novel approach that integrates feature representations extracted by discriminative models into conditional diffusion-based generative models. Experimental results demonstrate that our method outperforms diffusion models alone in vocal enhancement tasks while exhibiting better generalization ability compared to the discrminative \ac{USS} model. Additionally, our approach maintains robust speech enhancement performance in typical noisy environments. Through experiments 
we show that our method effectively combines the high vocal quality of \ac{USS} discriminative models with the strong generalization capabilities of diffusion models. In future work, we plan to explore combining discriminative models with other generative models and experimenting with alternative integration architectures.

\section*{Acknowledgment}
This project is funded by the NSFC Young Scientists Fund (Category C) under Grant No. 62501512.

%
%
%
\bibliographystyle{splncs04}
\bibliography{refs}

\begin{thebibliography}{10}
\providecommand{\url}[1]{\texttt{#1}}
\providecommand{\urlprefix}{URL }
\providecommand{\doi}[1]{https://doi.org/#1}

\bibitem{anderson1982reverse}
Anderson, B.: Reverse-time diffusion equation models. Stochastic Processes and
  their Applications  \textbf{12},  313--326 (1982)

\bibitem{bralios2023latent}
Bralios, D., Tzinis, E., Wichern, G., Smaragdis, P., Roux, J.L.: Latent
  iterative refinement for modular source separation. In: Proc. ICASSP. Rhodes
  Island (2023)

\bibitem{brock2018large}
Brock, A., Donahue, J., Simonyan, K.: Large scale {GAN} training for high
  fidelity natural image synthesis. In: Proc. ICLR. Vancouver (2018)

\bibitem{chao2024mamba}
Chao, R., Cheng, W.H., La~Quatra, M., Siniscalchi, S.M., Yang, C.H.H., Fu,
  S.W., Tsao, Y.: An investigation of incorporating mamba for speech
  enhancement. arXiv preprint arXiv:2405.06573  (2024)

\bibitem{defossez2019demucs}
D{\'e}fossez, A., Usunier, N., Bottou, L., Bach, F.: Demucs: {D}eep extractor
  for music sources with extra unlabeled data remixed. arXiv preprint
  arXiv:1909.01174  (2019)

\bibitem{defossez2019music}
D{\'e}fossez, A., Usunier, N., Bottou, L., Bach, F.: Music source separation in
  the waveform domain. arXiv preprint arXiv:1911.13254  (2019)

\bibitem{fang2021variational}
Fang, H., Carbajal, G., Wermter, S., Gerkmann, T.: Variational autoencoder for
  speech enhancement with a noise-aware encoder. In: Proc. ICASSP. Toronto
  (2021)

\bibitem{fonseca2021selfsupervised}
Fonseca, E., Jansen, A., Ellis, D.P.W., Wisdom, S., Tagliasacchi, M., Hershey,
  J.R., Plakal, M., Hershey, S., Moore, R.C., Serra, X.: Self-supervised
  learning from automatically separated sound scenes. In: Proc. WASPAA. New
  Paltz (2021)

\bibitem{gemmeke2017audio}
Gemmeke, J., Ellis, D., Freedman, D., Jansen, A., Lawrence, W., Moore, R.,
  Plakal, M., Ritter, M.: {AudioSet}: {A}n ontology and human-labeled dataset
  for audio events. In: Proc. ICASSP. New Orleans (2017)

\bibitem{gerkmann2015phase}
Gerkmann, T., Krawczyk-Becker, M., Le~Roux, J.: Phase processing for
  single-channel speech enhancement:{ H}istory and recent advances. IEEE Signal
  Processing Magazine  \textbf{32},  55--66 (2015)

\bibitem{goodfellow2014generative}
Goodfellow, I., Pouget-Abadie, J., Mirza, M., Xu, B., Warde-Farley, D., Ozair,
  S., Courville, A., Bengio, Y.: Generative adversarial nets. In: Proc. NIPS.
  Montreal (2014)

\bibitem{helwani2024soundsourceseparationusing}
Helwani, K., Togami, M., Smaragdis, P., Goodwin, M.M.: Sound source separation
  using latent variational block-wise disentanglement. arXiv preprint
  arXiv:2402.06683  (2024)

\bibitem{hendriks2022dft}
Hendriks, R., Gerkmann, T., Jensen, J.: {DFT}-domain based single-microphone
  noise reduction for speech enhancement. Springer Nature (2022)

\bibitem{ho2020denoising}
Ho, J., Jain, A., Abbeel, P.: Denoising diffusion probabilistic models. In:
  Proc. NeurIPS (2020)

\bibitem{jansson2017singing}
Jansson, A., Humphrey, E., Montecchio, N., Bittner, R., Kumar, A., Weyde, T.:
  Singing voice separation with deep {U-Net} convolutional networks. In: Proc.
  ISMIR. Suzhou (2017)

\bibitem{jensen2016algorithm}
Jensen, J., Taal, C.: An algorithm for predicting the intelligibility of speech
  masked by modulated noise maskers. IEEE Transactions on Audio, Speech, and
  Language Processing  \textbf{24},  2009--2022 (2016)

\bibitem{jukic2024schr}
Jukic, A., Korostik, R., Balam, J., Ginsburg, B.: Schr{\"o}dinger bridge for
  generative speech enhancement. arXiv preprint arXiv:2407.16074  (2024)

\bibitem{karras2020analyzing}
Karras, T., Laine, S., Aittala, M., Hellsten, J., Lehtinen, J., Aila, T.:
  Analyzing and improving the image quality of {StyleGAN}. In: Proc. CVPR.
  Seattle (2020)

\bibitem{kim2021multi}
Kim, H., Yoon, J., Cheon, S., Kang, W., Kim, N.: A multi-resolution approach to
  {GAN}-based speech enhancement. Applied Sciences  \textbf{11}, ~721 (2021)

\bibitem{koizumi2020speechenhancement}
Koizumi, Y., Yatabe, K., Delcroix, M., Masuyama, Y., Takeuchi, D.: Speech
  enhancement using self-adaptation and multi-head self-attention. In: Proc.
  ICASSP. Virtual Barcelona (2020)

\bibitem{kong2020panns}
Kong, Q., Cao, Y., Iqbal, T., Wang, Y., Wang, W., Plumbley, M.: {PANNs}:
  {L}arge-scale pretrained audio neural networks for audio pattern recognition.
  IEEE/ACM Transactions on Audio, Speech, and Language Processing  \textbf{28},
   2880--2894 (2020)

\bibitem{kong2021decoupling}
Kong, Q., Cao, Y., Liu, H., Choi, K., Wang, Y.: Decoupling magnitude and phase
  estimation with deep {ResUNet} for music source separation. In: Proc. ISMIR
  (2021)

\bibitem{kong2023universal}
Kong, Q., Chen, K., Liu, H., Du, X., Berg-Kirkpatrick, T., Dubnov, S.,
  Plumbley, M.: Universal source separation with weakly labelled data. arXiv
  preprint arXiv:2305.07447  (2023)

\bibitem{le_roux2019sdr}
Le~Roux, J., Wisdom, S., Erdogan, H., Hershey, J.: {SDR}--{H}alf-baked or well
  done? In: Proc. ICASSP. Brighton (2019)

\bibitem{lemercier2023diffusion}
Lemercier, J.M., Richter, J., Welker, S., Gerkmann, T.: Analysing
  diffusion-based generative approaches versus discriminative approaches for
  speech restoration. In: Proc. ICASSP. Rhodes Island (2023)

\bibitem{li2024diffusionbasedgenerativemodel}
Li, C., Cornell, S., Watanabe, S., Qian, Y.: Diffusion-based generative
  modeling with discriminative guidance for streamable speech enhancement.
  arXiv preprint arXiv:2406.13471  (2024)

\bibitem{sepanddiff}
Lutati, S., Nachmani, E., Wolf, L.: Separate and diffuse: {U}sing a pretrained
  diffusion model for better source separation. In: Proc. ICLR. Vienna (2024)

\bibitem{opensourceseparation:book}
Manilow, E., Seetharman, P., Salamon, J.: Open Source Tools \& Data for Music
  Source Separation. Source Separation GitHub Tutorial (2020),
  \url{https://source-separation.github.io/tutorial}

\bibitem{plaja2022diffusion}
Plaja-Roglans, G., Miron, M., Serra, X.: A diffusion-inspired training strategy
  for singing voice extraction in the waveform domain. In: Proc. ISMIR.
  Bengaluru (2022)

\bibitem{rafii2017musdb18}
Rafii, Z., Liutkus, A., St{\"o}ter, F.R., Mimilakis, S., Bittner, R.: The
  {MUSDB18} corpus for music separation (2017),
  \url{https://doi.org/10.5281/zenodo.1117372}

\bibitem{rezende2015variational}
Rezende, D., Mohamed, S.: Variational inference with normalizing flows. In:
  Proc. ICML. Lille (2015)

\bibitem{richter2023speech}
Richter, J., Welker, S., Lemercier, J.M., Lay, B., Gerkmann, T.: Speech
  enhancement and dereverberation with diffusion-based generative models.
  IEEE/ACM Transactions on Audio, Speech, and Language Processing  \textbf{31},
   2351--2364 (2023)

\bibitem{rix2001perceptual}
Rix, A., Beerends, J., Hollier, M., Hekstra, A.: Perceptual evaluation of
  speech quality ({PESQ}) - {A} new method for speech quality assessment of
  telephone networks and codecs. In: Proc. ICASSP. Salt Lake City (2001)

\bibitem{roweis2000one}
Roweis, S.: One microphone source separation. Advances in neural information
  processing systems  \textbf{13} (2000)

\bibitem{shi2024ensemble}
Shi, H., Kamo, N., Delcroix, M., Nakatani, T., Araki, S.: Ensemble inference
  for diffusion model-based speech enhancement. In: Proc. ICASSP. Seoul (2024)

\bibitem{song2019generative}
Song, Y., Ermon, S.: Generative modeling by estimating gradients of the data
  distribution. In: Proc. NeurIPS. Vancouver (2019)

\bibitem{song2020score}
Song, Y., Sohl-Dickstein, J., Kingma, D., Kumar, A., Ermon, S., Poole, B.:
  Score-based generative modeling through stochastic differential equations.
  In: Proc. ICLR (2020)

\bibitem{stoller2018wave}
Stoller, D., Ewert, S., Dixon, S.: {Wave-U-Net}: {A} multi-scale neural network
  for end-to-end audio source separation. In: Proc. ISMIR. Paris (2018)

\bibitem{sarkka2019applied}
Särkkä, S., Solin, A.: Applied Stochastic Differential Equations. Cambridge
  University Press (2019)

\bibitem{thiemann2013diverse}
Thiemann, J., Ito, N., Vincent, E.: The diverse environments multi-channel
  acoustic noise database ({DEMAND}): {A} database of multichannel
  environmental noise recordings. In: Proc. of Meetings on Acoustics. San
  Francisco (2013)

\bibitem{tzinis2020sudo}
Tzinis, E., Wang, Z., Smaragdis, P.: Sudo rm-rf: {E}fficient networks for
  universal audio source separation. In: Proc. MLSP. Espoo (2020)

\bibitem{Vaswani2017AttentionIA}
Vaswani, A., Shazeer, N., Parmar, N., Uszkoreit, J., Jones, L., Gomez, A.,
  Kaiser, L., Polosukhin, I.: Attention is all you need. In: Proc. NIPS. Long
  Beach (2017)

\bibitem{vincent2011connection}
Vincent, P.: A connection between score matching and denoising autoencoders.
  Neural Computation  \textbf{23},  1661--1674 (2011)

\bibitem{wang2024diffusion}
Wang, S., Liu, S., Harper, A., Kendrick, P., Salzmann, M., Cernak, M.:
  Diffusion-based speech enhancement with schr\"{o}dinger bridge and symmetric
  noise schedule. arXiv preprint arXiv:2409.05116  (2024)

\bibitem{wang2023framework}
Wang, Z., Giri, R., Shah, D., Valin, J.M., Goodwin, M., Smaragdis, P.: A
  framework for unified real-time personalized and non-personalized speech
  enhancement. In: Proc. ICASSP. Rhodes Island (2023)

\bibitem{welker2022speech}
Welker, S., Richter, J., Gerkmann, T.: Speech enhancement with score-based
  generative models in the complex {STFT} domain. In: Proc. Interspeech.
  Incheon (2022)

\bibitem{zhang2023towarduniversal}
Zhang, W., Saijo, K., Wang, Z.Q., Watanabe, S., Qian, Y.: Toward universal
  speech enhancement for diverse input conditions. In: Proc. ASRU. Taipei
  (2023)

\end{thebibliography}
%




\end{document}